\documentclass[a4paper]{jpconf}
\usepackage{graphicx}

\begin{document}
\title{The CDEX Dark Matter Program at the
China Jinping Underground Laboratory}

\author{
Qian Yue\footnote{Corresponding Author},
Kejun Kang,
Jianming Li
}
\address{
Department of Engineering Physics, Tsinghua University,
Beijing 100084.
}
\ead{$^1$yueq@mail.tsinghua.edu.cn}

\author{Henry T. Wong$^1$}
\address{
Institute of Physics, Academia Sinica, Taipei 11529.
}
\ead{$^1$htwong@phys.sinica.edu.tw}

\begin{abstract}

The China Jinping Underground Laboratory (CJPL)
is a new facility for conducting low event-rate experiments.
We present an overview of CJPL and 
the CDEX Dark Matter program
based on germanium detectors with sub-keV
sensitivities.
The achieved results, status as well as 
the R\&D and technology acquisition efforts 
towards a ton-scale experiment 
are reported.

\end{abstract}

\section{China Jinping Underground Laboratory}

The China Jinping Underground Laboratory (CJPL)\cite{cjpl}
is located in Sichuan, China, and
was inaugurated in December 2012. 
With a rock overburden of about 2400~meter,
it is the deepest operating underground laboratory in
the world. 
The muon flux is measured to be
$( 2.0 \pm 0.4 ) \times 10^{-10} \rm{cm^{-2} s^{-1}}$\cite{cjpl-cosmic}, 
suppressed from the sea-level flux by a factor of $10^{-8}$.
The drive-in tunnel access can greatly facilitate the
deployment of big experiments and large teams.
Supporting infrastructures of catering and accommodation, 
as well as office and workshop spaces, 
already exist.

\begin{figure}
\begin{center}
\includegraphics[width=10cm]{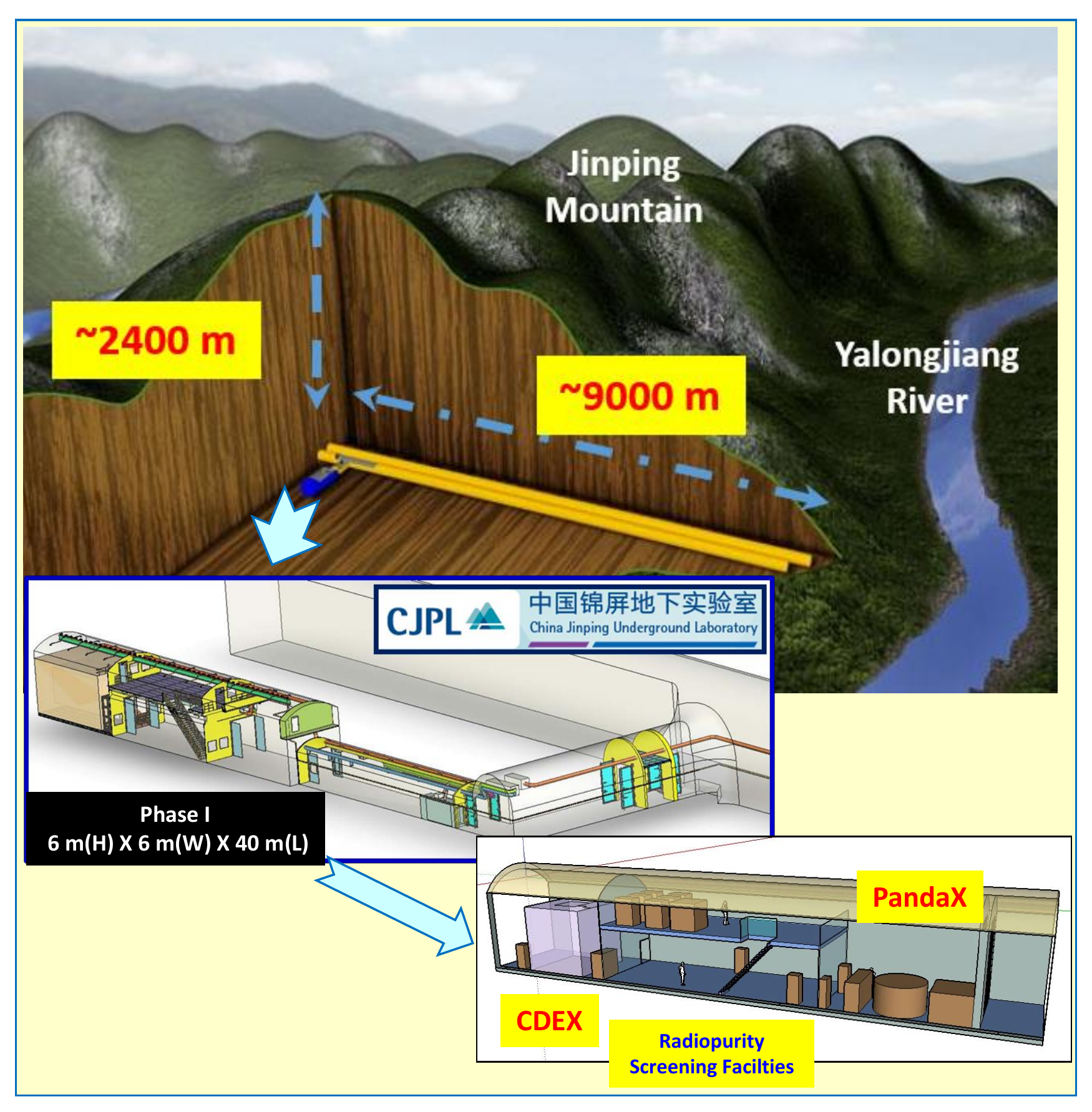}
\caption{\label{fig::cjpl1} 
Schematic diagram of CJPL~Phase-I inaugurated in 2012, 
showing the space allocation
to the CDEX and PandaX Dark Matter experiments, as well as to
the radiopurity screening facilities.
}
\end{center}
\end{figure}

As depicted schematically in Figure~\ref{fig::cjpl1},
the completed CJPL~Phase-I consist of a laboratory hall
of dimension 6~m(W)$\times$ 6~m(H)$\times$40~m(L).
This space is currently used by the CDEX\cite{cdex} 
and PandaX\cite{pandax} dark matter
experiments, as well as for a general purpose low radiopurity
screening facility.

\begin{figure}
\begin{center}
{\bf (a)}\\
\includegraphics[width=10cm]{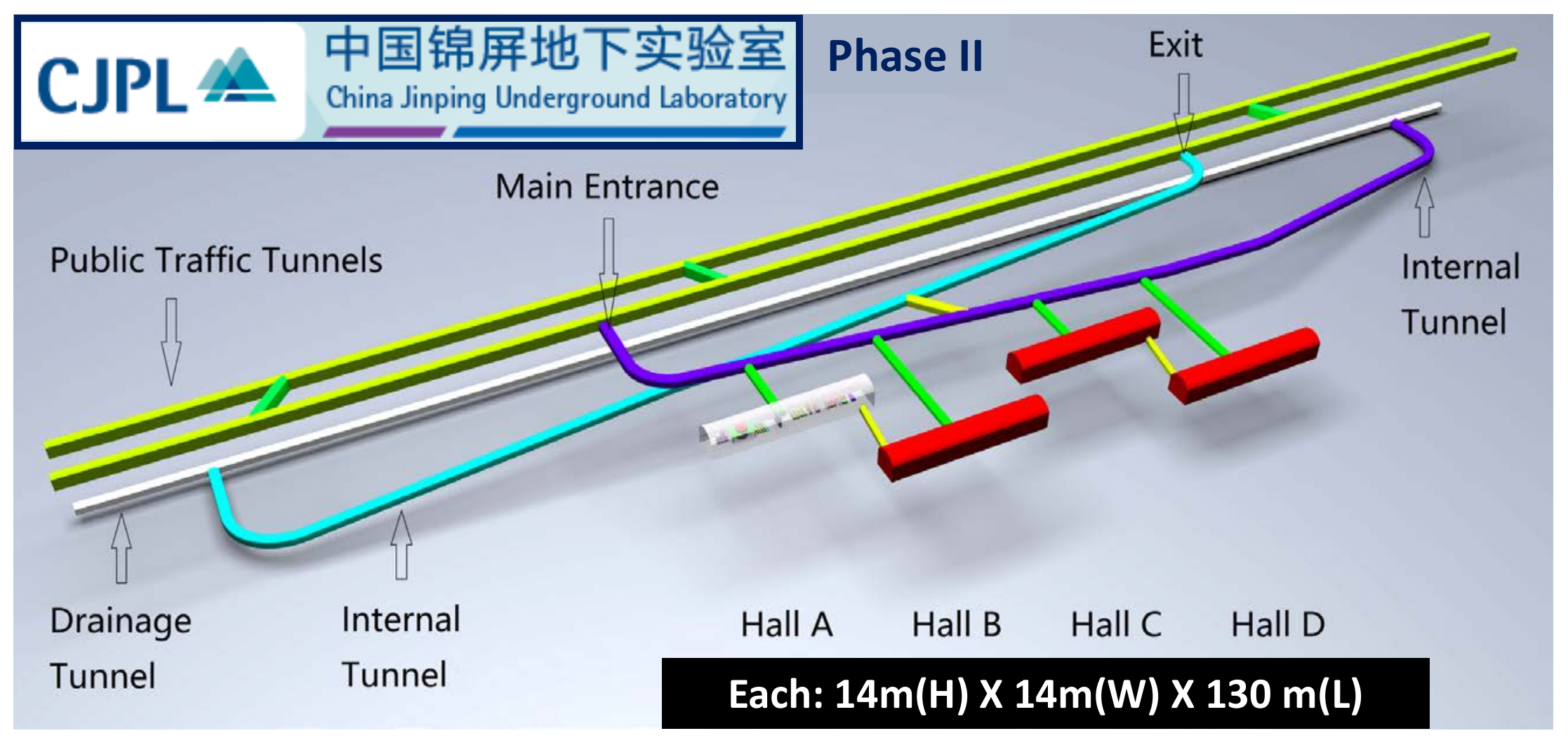}\\
{\bf (b)}\\
\includegraphics[width=10cm]{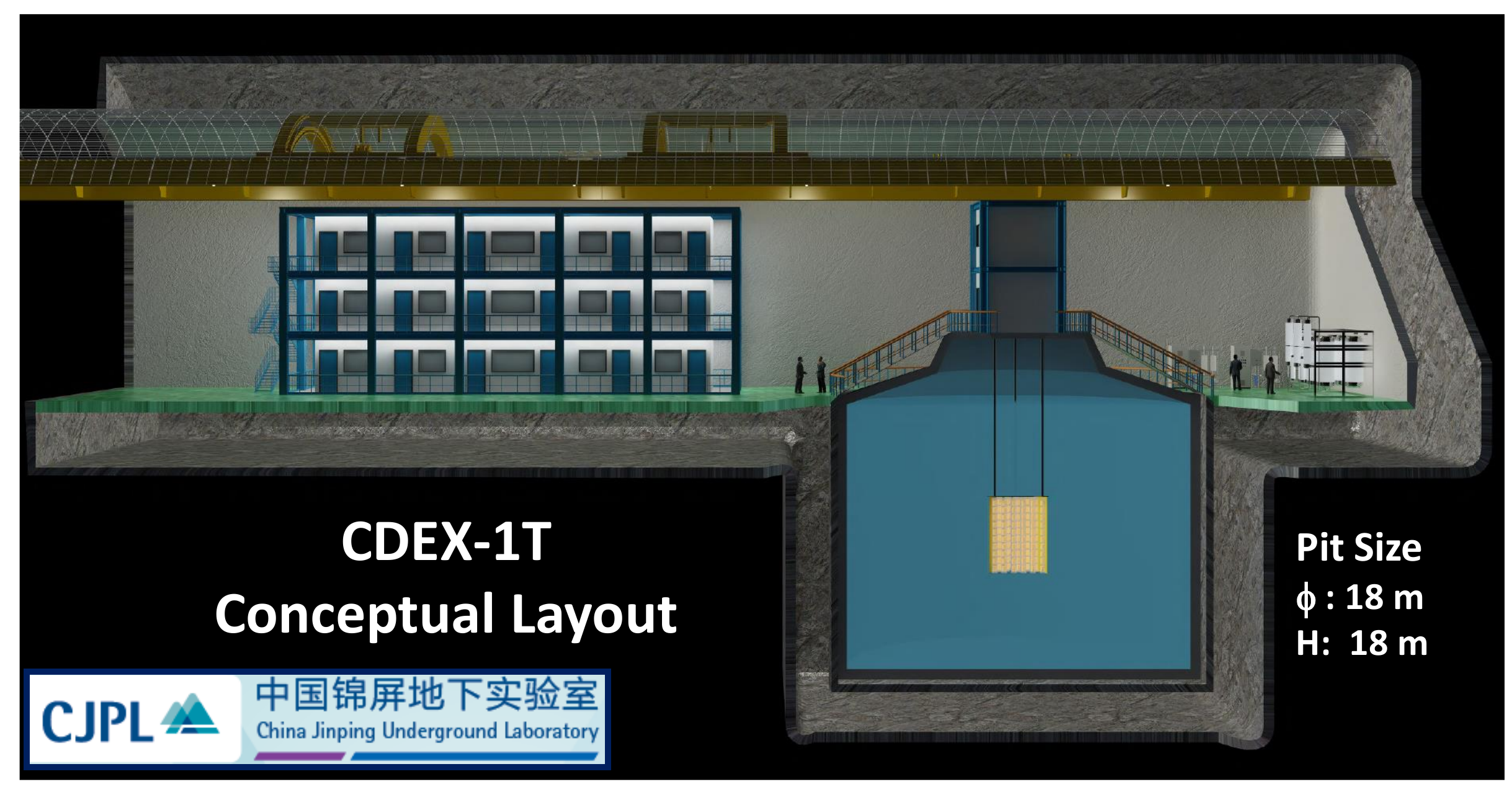}
\caption{\label{fig::cjpl2} 
(a) Schematic diagram of CJPL~Phase-II scheduled to complete
by early 2017. 
(b) Conceptual configuration of a future
CDEX-1T experiment at CJPL~Phase-II.
}
\end{center}
\end{figure}

Additional laboratory space for CJPL~Phase-II, 
located about 500~m from the Phase-I site,
is currently under construction.
Upon the scheduled completion by early 2017,
it will consist of four halls each with dimension
14~m(W)$\times$14~m(H)$\times$130~m(L).
The tunnel layout is as displayed in Figure\ref{fig::cjpl2}a.

\section{CDEX Dark Matter Program}

About one-quarter of the energy density of the Universe 
can be attributed to cold dark matter~\cite{rpp-dm},
whose nature and properties are unknown. 
Weakly interacting massive particles (WIMPs denoted by $\chi$) 
are its leading candidates. The WIMPs interact with matter 
pre-dominantly via elastic scattering with the nucleus: 
$\chi + N \rightarrow \chi + N$ . 
The unique advantages of CJPL make it an ideal location
to perform experiments on dark matter searches.

Germanium detectors sensitive to sub-keV recoil energy were 
identified and demonstrated as possible means to probe the 
``light'' WIMPs with mass range
${\rm 1~GeV < {m_{\chi}} < 10~GeV}$\cite{ge-lightwimp}. 
This inspired development of p-type point-contact 
germanium detectors (pPCGe) with modular mass of kg-scale\cite{ppcge}, 
followed by various experimental efforts\cite{cogent,texono,malbek}.
The scientific theme of CDEX (China Dark matter EXperiment)\cite{cdex} 
is to pursue studies of light WIMPs with pPCGe.
It is one of the two founding experimental programs at CJPL.

\subsection{First Generation CDEX Experiments}

\begin{figure}
\begin{center}
\includegraphics[width=10cm]{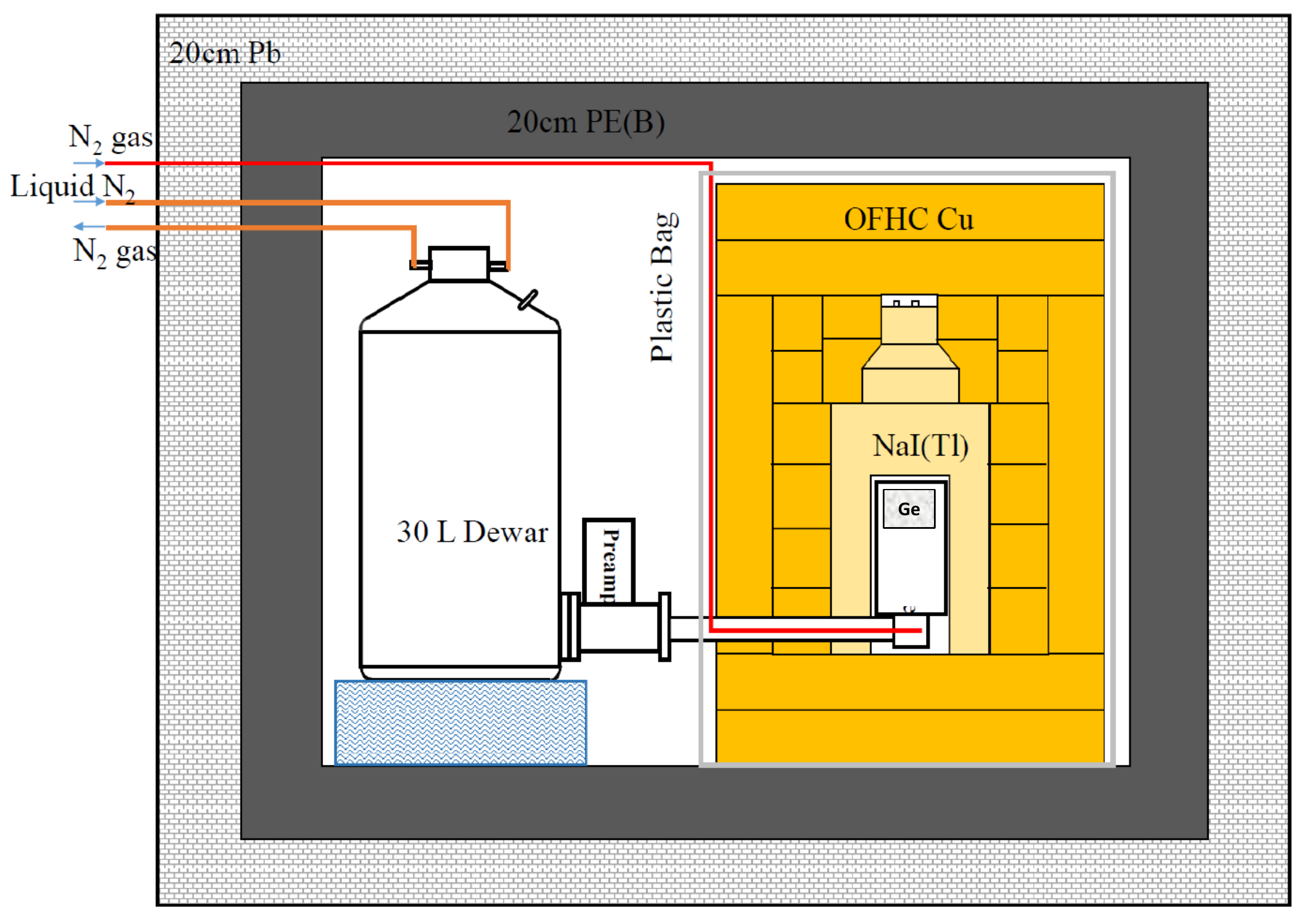}
\caption{\label{fig::cdex1-design}
Schematic diagram of the baseline design of the
CDEX-0 and CDEX-1 experiments, 
using single-element pPCGe detector
enclosed by NaI(Tl) crystal scintillator and
passive shieldings.
}
\end{center}
\end{figure}

As depicted in Figure~\ref{fig::cdex1-design},
the first-generation experiments adopted 
a baseline design\cite{texono} 
of single-element ``1-kg mass scale''
pPCGe enclosed by NaI(Tl) crystal 
scintillator as anti-Compton detectors. 
These active detectors
are further surrounded by passive shieldings of OFHC copper,
boron-loaded polyethylene (PE(B)) and lead, 
while the detector volume is
purged by dry nitrogen to suppress radon contamination.
 
\begin{figure}
\begin{center}
\includegraphics[width=12cm]{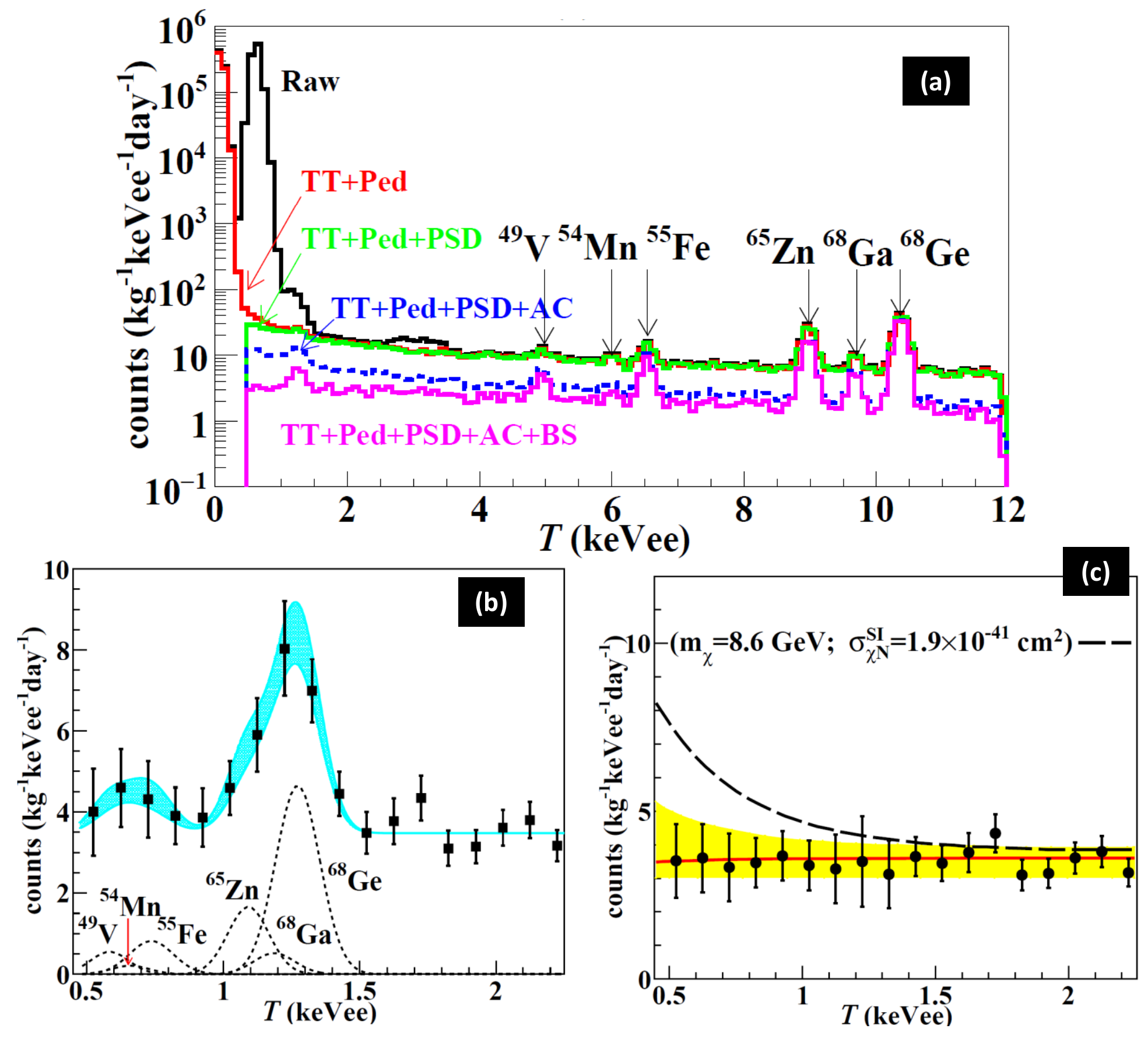}
\caption{\label{fig::cdex1-spectra}
(a)
Background spectra of the CDEX-1 measurement 
at their various stages of selection:
basic cuts (TT+Ped+PSD), Anti-Compton (AC) and Bulk (BS) events.
(b)
All events can be accounted for with 
the known background channels $-$ L-shell X-rays and flat background
due to ambient high energy $\gamma$-rays. 
(c)
Examples of excluded $\chi$N recoil spectra are superimposed.
}
\end{center}
\end{figure}

\begin{figure}
\begin{center}
\includegraphics[width=12cm]{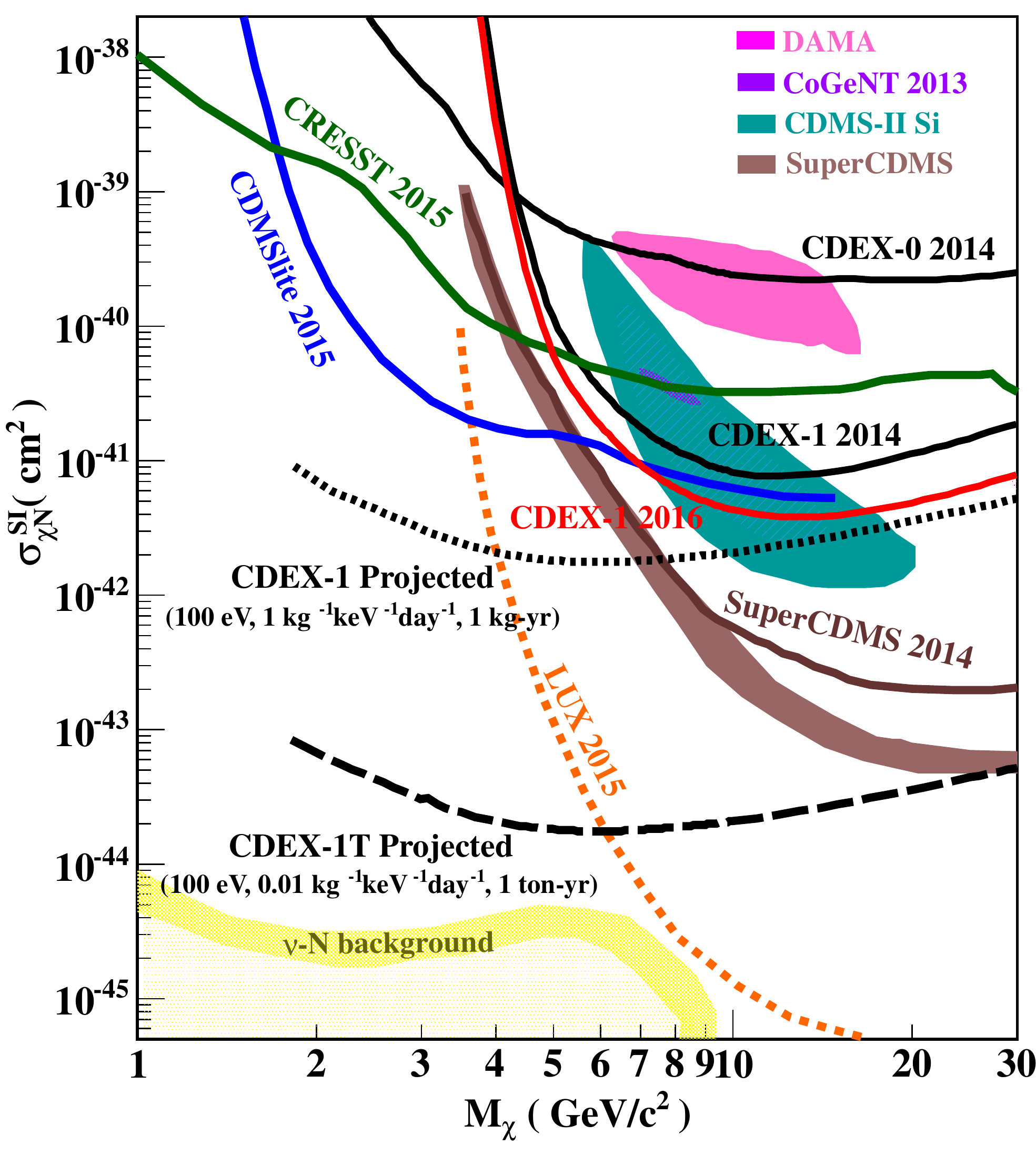}
\caption{\label{fig::cdex-explot}
Exclusion regions derived from the CDEX-0 and CDEX-1 
experiments, and comparison with other benchmark results.
Projected sensitivities of the current detectors 
and future projects are superimposed.
}
\end{center}
\end{figure}

The pilot CDEX-0 measurement is based on 
a 20~g prototype Ge detector
at 177~$\rm(eV_{ee})$ threshold with an exposure
of 0.784~kg-days\cite{cdex0}.
The CDEX-1 experiment
adopts a pPCGe detector of mass 1~kg.
The first results are based on 
an analysis threshold of 475~$\rm{eV_{ee}}$ 
with an exposure of 53.9~kg-days\cite{cdex1}.
After suppression of the anomalous surface background events
and measuring their signal efficiencies and background 
leakage factors with calibration data\cite{bsel}, all residual events
can be accounted for by known background models.
The updated results with 335.6~kg-days\cite{cdex1}
of exposure are displayed in
Figure~\ref{fig::cdex1-spectra}. 
Dark Matter constraints on $\chi$N spin-independent
cross-sections were derived for both data set, 
and are displayed in Figure~\ref{fig::cdex-explot},
together with other selected benchmark results~\cite{taup15}. 
In particular, the allowed region from 
the CoGeNT\cite{cogent} experiment 
is probed and excluded with the CDEX-1 results.

Analysis is currently performed on CDEX-1 data set with year-long
exposure.
Annual modulation effects as well as other physics channels 
are being studied. New data is also taken with an upgraded pPCGe 
with lower threshold.

\subsection{Current Efforts and Future Goals}

The long-term goal of the CDEX program will be
a ton-scale germanium experiment (CDEX-1T)
at CJPL for the
searches of dark matter and of neutrinoless
double beta decay ($0 \nu \beta \beta$)\cite{0nubb}.
A pit of diameter 18~m and height 18~m will be built
at one of the halls of CJPL-Phase~II to house such
an experiment, as illustrated in Figure~\ref{fig::cjpl2}b.

Towards this ends, the ``CDEX-10'' prototype  
has been constructed with detectors in array structure
having a target mass at the 10-kg range.
This would provide a platform to study
the many issues of scaling up in detector mass and in 
improvement of background and threshold.
The detector array is shielded and cooled by a cryogenic liquid.
Liquid nitrogen is being used, while liquid argon is 
a future option to investigate, which may offer the
additional potential benefits of 
an active shielding as anti-Compton detector.

In addition, various crucial technology acquisition projects
are pursued, which would make a ton-scale germanium experiment 
realistic and feasible. 
These include: 
\begin{enumerate}
\item detector grade germanium crystal growth;
\item germanium detector fabrication;
\item isotopic enrichment of $^{76}$Ge for $0 \nu \beta \beta$;
\item production of electro-formed copper, eventually underground at CJPL.
\end{enumerate}

The first detector fabricated by the Collaboration from 
commercial crystal that matches expected performance 
will be installed at CJPL in 2016. It allows control of
assembly materials placed at its vicinity, known to be
the dominant source of radioactive background, as well
as efficient testing of novel electronics and readout
schemes. The benchmark would be to perform light WIMP searches
with germanium detectors with ``$0 \nu \beta \beta$-grade''
background control. 
This configuration would provide the
first observation (or stringent upper bounds) 
of the potential cosmogenic tritium contaminations in 
germanium detectors, from which 
the strategies to suppress such background
can be explored.

The projected $\chi$N sensitivity for CDEX-1T
is shown in Figure~\ref{fig::cdex-explot}, 
taking a realistic minimal
surface exposure of six months. 
The goal for $0 \nu \beta \beta$ will be to
achieve sensitivities covering completely
the inverted neutrino mass hierarchy.

\end{document}